# Realization of the Haldane Chern insulator in a moiré lattice

Wenjin Zhao[1], Kaifei Kang[2], Lizhong Li[2], Charles Tschirhart[3], Evgeny Redekop[3], Kenji Watanabe[4], Takashi Taniguchi[4], Andrea Young[3], Jie Shan[1,2,5*], Kin Fai Mak[1,2,5*]

[1]Kavli Institute at Cornell for Nanoscale Science, Ithaca, NY, USA
[2]School of Applied and Engineering Physics, Cornell University, Ithaca, NY, USA
[3]Department of Physics, University of California at Santa Barbara, Santa Barbara, CA, USA
[4]National Institute for Materials Science, 1-1 Namiki, 305-0044 Tsukuba, Japan
[5]Laboratory of Atomic and Solid State Physics, Cornell University, Ithaca, NY, USA

Email: jie.shan@cornell.edu; kinfai.mak@cornell.edu

**The Chern insulator displays a quantized Hall effect without Landau levels [1–7]. In a landmark paper in 1988, Haldane showed that a Chern insulator could be realized through complex next-nearest-neighbor hopping in a honeycomb lattice [1]. Despite its profound impact on the field of topological physics [8,9] and recent implementation in cold-atom experiments [10], the Haldane model has remained elusive in solid-state materials. Here, we report the experimental realization of a Haldane Chern insulator in AB-stacked $MoTe_2/WSe_2$ moiré bilayers, which form a honeycomb moiré lattice with two sublattices residing in different layers [7,11–15]. We show that the moiré bilayer filled with two charge particles per unit cell is a quantum spin Hall (QSH) insulator with a tunable charge gap. Under a small out-of-plane magnetic field, it becomes a Chern insulator with Chern number $c = 1$ from magneto-transport studies. The results are qualitatively captured by a generalized Kane-Mele tight-binding Hamiltonian [11–17]. The Zeeman field splits the QSH insulator into two halves of opposite valley--one with a positive and the other a negative moiré band gap. Our study highlights the unique potential of semiconductor moiré materials in engineering topological lattice Hamiltonians [18–20].**

## Main

When a two-dimensional (2D) electron gas is subjected to high magnetic fields and low temperatures, the spectrum of electron energy levels is split into Landau levels, and quantized Hall conductance can be observed [21]. The quest for a quantum Hall state without Landau levels or even external magnetic fields [1], that is, a Chern or quantum anomalous Hall (QAH) insulator, is driven by both fundamental and technological interests. To date, only a handful of materials have exhibited the QAH effect, including magnetic topological insulators $(Bi,Sb)_2Te_3$ and $MnBi_2Te_4$, and graphene and transition metal dichalcogenide (TMD) moiré materials at half-band filling [2–7]. A distinct proposal for realizing a Chern insulator is to 'split' through magnetic interactions a QSH insulator, which can be viewed as two time-reversal copies of the Chern insulator [22] (Fig. 1). After the time-reversal symmetry is broken, one spin species can have a positive band gap corresponding to a normal band insulator, while the other has a negative band gap with inverted bands, giving rise to a topologically nontrivial insulator with quantized Hall conductance. The idea hinges on specific band-dependent magnetic interactions and Zeeman energies exceeding the QSH band gap, which are exceedingly difficult to satisfy in known QSH materials [23–26].

Here we overcome both challenges in AB-stacked (60-degree-aligned) MoTe$_2$/WSe$_2$ moiré bilayers, a new QSH insulator with a tunable charge gap by an out-of-plane electric field and a unique spin-valley-layer locked band structure [7,11–15] (Fig. 1). We show that the application of an out-of-plane magnetic field on the order of 1 T transforms the material into a Chern insulator without forming Landau levels. The system in principle also allows a Chern insulator state without an external magnetic field, for instance, by proximity coupling to a layered magnetic insulator [27,28]. Our results provide the basis for future such studies.

**AB-stacked MoTe$_2$/WSe$_2$ moiré bilayer**
The Wannier orbital of the first (second) moiré valence band in the TMD moiré bilayer is centered at the MM (XX) stacking point in the MoTe$_2$ (WSe$_2$) layer (M = Mo, W; X = Te, Se) [11,14]. They constitute the two sublattices of a honeycomb lattice (Fig. 1a). The electronic spin and valley degrees of freedom are locked because of the Ising spin-orbit interaction and time reversal invariance in each monolayer [18,19]. The spins of the two bands are anti-aligned in AB-stacked bilayers. The system realizes a generalized Kane-Mele tight-binding Hamiltonian [11–17] (Methods). Particularly, the next-nearest-neighbor (or intralayer) hopping has a complex value because of time reversal symmetry breaking in each (K or K') valley of the monolayers. The lattice relaxation and corrugation in the moiré structure break mirror symmetry, and enable the nearest-neighbor (or interlayer) hopping, which is otherwise spin-forbidden. An out-of-plane electric field, $E$, can tune the sublattice potential difference and induce band inversion.

At doping density of two charge particles per moiré unit cell ($\nu = 2$), after band inversion, the system is expected to transition from a normal band insulator to a QSH insulator with an insulating bulk and two counter-propagating edge states carrying opposite spin. The initial experimental evidence of such a topological phase transition has been recently reported [7]. Upon the application of an out-of-plane magnetic field, $B$, the spin-valley Zeeman effect can reduce the moiré band gap in one valley and eventually un-invert the bands, while enlarge the gap in the other (Fig. 1b-d). As a result, one of the edge states penetrates deeper into the bulk and eventually disappears, while the other bounds more strongly to the edge [22]. A Haldane Chern insulator thus emerges. Furthermore, because the band gap is continuously tunable by $E$, the system in principle could be tuned to the quantum critical point with linear band crossing; an infinitesimal magnetic field would be sufficient to induce the Haldane Chern insulator. The intuitive picture is supported by the tight-binding calculations including the Zeeman energy for AB-stacked MoTe$_2$/WSe$_2$ moiré bilayers (Methods). This mechanism of realizing a Haldane Chern insulator is unique to the AB-stacked bilayers. It is not expected in AA-stacked bilayers, which have strongly hybridized bands from interlayer hopping [29], and in graphene moiré systems [30–33].

**Quantum spin Hall insulator**
We perform charge transport studies on multiple AB-stacked MoTe$_2$/WSe$_2$ moiré bilayers using a dual-gated Hall bar structure. The two gate voltages independently control both the sample doping density and the out-of-plane electric field. The moiré unit cell density ($n_{\text{M}} \approx 5 \times 10^{12}$ cm$^{-2}$) is primarily determined by the large lattice mismatch between the two TMDs. The effect of twist angle disorder is reduced compared to that in moiré homobilayers [34,35]. The device boundaries are electrostatically defined by the top gate, which are atomically smooth and significantly suppress back scattering of the helical edge states (see Methods for details on device fabrication and characterization).

In the absence of a magnetic field, the Hall conductance vanishes by time reversal symmetry in a QSH insulator. We measure the longitudinal conductance using the geometry shown in the inset of Fig. 2c. The device is current-biased along the long axis of the Hall bar, and the voltage drop is measured between two adjacent electrodes separated by 2.5 μm on the same side of the Hall bar. If charge transport is dominated by the helical edge states, the longitudinal four-terminal resistance, $R_{\text{xx}}$, is expected to take a quantized value of $\left(\frac{1}{2}\right)\frac{h}{e^2}$, where $h$ and $e$ denote the Planck's constant and the elementary charge, respectively. Figure 2a shows $R_{\text{xx}}$ of device 1 as a function of the top and bottom gate voltages, $V_{\text{tg}}$ and $V_{\text{bg}}$, at 0.33 K. The two dashed lines identify, respectively, the axes along which the doping density is fixed at $\nu = 2$ and the electric field at $E_{\text{c}} = 0.425$ V/nm; and the arrows show the direction of increasing $E$ or $\nu$.

Figure 2c shows the electric-field dependence of $R_{\text{xx}}$ at $\nu = 2$ and temperature $T$ ranging from 0.33 K to 30 K. The resistance shows a minimum near $E_{\text{c}}$ and distinct behaviors on two sides of $E_{\text{c}}$. Below $E_{\text{c}}$, $R_{\text{xx}}$ decreases as $E$ approaches $E_{\text{c}}$; and at a fixed field, $R_{\text{xx}}$ diverges as the temperature decreases. This is a typical response of an insulator with a diminishing band gap towards $E_{\text{c}}$. Above $E_{\text{c}}$, $R_{\text{xx}}$ plateaus, and the value saturates around 15 kΩ $\approx 1.16\ h/2e^2$ at 0.33 K. The nearly quantized $R_{\text{xx}}$ plateau suggests the emergence of a QSH insulator for $E > E_{\text{c}}$, where $E_{\text{c}}$ corresponds to the quantum critical point for band inversion. The presence of the helical edge states is also consistent with the observed nonlocal transport (Extended Data Fig. 1). The result is further supported by a comparison to the bulk resistance, $R_{\text{xx\_bulk}}$, in Fig. 2d. In this case, the device is biased along the short axis of the Hall bar, and the voltage drop is measured by an adjacent pair of electrodes along the same direction. An additional contact on each side of the probe region is grounded to avoid the edge current. Again, we observe a resistance minimum at $E_{\text{c}}$, but the response on both sides of the minimum is one of an insulator.

We compare the temperature dependence of $R_{\text{xx}}$ and $R_{\text{xx\_bulk}}$ after band inversion at $E = 0.485$ V/nm in Fig. 2b. At high temperatures or small $T^{-1}$, both show an activation behavior. We estimate the gap size to be 2-3 meV from $R_{\text{xx\_bulk}}$. At low temperatures, the bulk states are largely localized. The transport is dominated by the edge states, which results in nearly quantized $R_{\text{xx}}$. But full quantization is prohibited by residual hopping transport in the bulk, as well as back scattering of the helical edge states. Nearly quantized edge conduction is observed in multiple devices. We also observe a dependence of the QSH phase space on the sample twist angle (Extended Data Fig. 2). The $R_{\text{xx}}$ plateaus over a wide range of $E$ in AB-bilayers with a small twist angle (device 1 and 3, ~ 2-3 degrees), and over a relatively narrow range near band inversion in nearly angle-aligned AB-bilayers (device 2 and 4).

Future studies are required to understand the systematic twist angle dependence, which is likely associated with the enhanced correlation effect in angle-aligned samples [18,34].

**Haldane Chern insulator**

We apply a perpendicular magnetic field and examine the transport properties. Figure 3 and 4 illustrate the result from device 2 at temperatures down to 10 mK (lattice temperature). Similar studies of device 1 down to 1.6 K are included in Extended Data Fig. 3. Figure 3a shows the Hall resistance $R_{xy}$ (top) and the longitudinal resistance $R_{xx}$ (bottom) as a function of gate voltages under $B = 2$ T at 10 mK. Results at other magnetic fields are included in Extended Data Fig. 4. The two dashed lines again identify the constant hole density $\nu = 2$ and critical electric field $E_c$ for band inversion. In contrast to measurements at higher temperatures ($> 1$ K), charge transport cannot be appropriately measured due to poor electrical contacts around $\nu = 2$ under small negative top gate voltages (black regions in Fig. 3a, b, see Methods for the contact design). The doping is electron-like or hole-like in the region below or above $\nu = 2$, which displays small $R_{xy}$ of different sign. The Landau levels are not observed. They start to emerge under a magnetic field above around 3 T.

The most striking feature in Fig. 3a is an unusually large $R_{xy}$ in the narrow region near $E_c$ and $\nu = 2$, which is correlated with a nearly vanishing $R_{xx}$. We examine the state in Fig. 3b by studying the magnetic-field dependence of $R_{xy}$ and $R_{xx}$ at constant electric-field $E_c$. The state can be identified either by a $R_{xy}$ maximum or a $R_{xx}$ minimum (denoted by dashed lines). It develops from $\nu = 2$ as the magnetic field increases, and shifts linearly with doping density and magnetic field. We determine the slope of the dashed lines to be $n_M \frac{d\nu}{dB} = c \frac{e}{h}$ with $c = 1.00 \pm 0.05$. Here $c$ is the Chern number of the occupied bands, and is related to the Hall resistance by the Streda relation [36], $R_{xy} = \frac{1}{c} \frac{h}{e^2}$. In the absence of Landau levels, the result suggests the emergence of a Chern insulator with Chern number $c = 1$.

We examine the magnetic field that is required to induce the Chern insulator. Figure 3c is the magnetic-field dependence of $R_{xy}$ and $R_{xx}$ along the dashed lines in Fig. 3b. The Hall resistance increases sharply from 0 at $B = 0$ T and plateaus between 1 T and 3 T. The plateau value ($\approx 25.4$ k$\Omega$) is within 2% of the quantized Hall resistance, $h/e^2$. Concurrently, $R_{xx}$ drops sharply with increasing magnetic field and remains small ($< 1$ k$\Omega$) between 1 T and 3 T. Hence, when the electric field is set near the quantum critical point for band inversion, a moderate magnetic field between 1 T and 3 T is sufficient to induce the Chern state.

The Chern insulator can be similarly induced by varying the magnetic field under a constant electric field both below and above $E_c$. We sample the phase diagram by varying $E$ under a constant $B$ (Extended Data Fig. 5). For each magnetic field, we observe a span of the electric field, $\Delta E$, near $E_c$ that hosts the Chern insulator state. The $E$-field span increases with field for small magnetic fields (Extended Data Fig. 5), which can be modeled by adding in quadrature a linear magnetic-field-dependent term and a constant ($\sim$ 2.2 mV/nm). The constant accounts for disorder broadening. Similar values have been observed in this type of devices ($\pm$ 2 mV/nm) [7,29].

Finally, we investigate the temperature dependence of the Chern insulator. We use the result under $B = 2$ T as an example. As temperature decreases, a sharp peak in the doping dependence of $R_{xy}$ (Fig. 4a), correlated with a sharp dip in $R_{xx}$ (Fig. 4b), emerges at $\nu =$ 2.01. The temperature dependence of the peak or dip values is summarized in Fig. 4c. The Hall resistance increases monotonically with decreasing temperature and becomes nearly quantized below 0.5 K. Simultaneously, the longitudinal resistance first increases as in an insulator, peaks around 3 K, and then drops below 1 k$\Omega$ below 0.5 K. The charge transport is therefore dominated by chiral edge transport below about 3 K, which provides an estimate of the charge gap size of the Chern state.

**Discussions and conclusion**
Our experimental results are fully consistent with the tight-binding calculations (Methods). The topological phase transition to a Haldane Chern insulator can be induced by varying the magnetic field from either a trivial moiré band insulator ($E < E_c$) or a QSH insulator ($E > E_c$) near band inversion. The phase boundaries of the Chern insulator expands linearly with magnetic field (Extended Data Fig. 6). This is governed by the competing Stark shift and the Zeeman shift, $(\Delta E)D \approx g\mu_B B$, where $g \approx 10$ is the g-factor of the TMD moiré valence band [37,38], $\mu_B$ is the Bohr magneton, and $D \approx 0.26$ e$\times$nm is the out-of-plane electric dipole of the moiré bilayer [29] (Methods). The termination of the Chern insulator at higher magnetic fields (for instance, $> 3$ T for $E = E_c$) is associated with the formation of Landau levels. The low-energy Landau levels disperse towards the Fermi level and introduce additional band inversions for $E > E_c$ (Methods). The Chern insulator thus disappears and reappears in an alternating sequence with increasing magnetic field (Extended Data Fig. 7).

We observe deviations from the ideal Hall quantization and dissipationless transport for the Chern insulator even at the lowest temperature in our experiment. This is originated from disorder broadening of energy levels ($\sim 0.5$ meV estimated from the onset magnetic field for the Landau levels) and of $E_c$ and $n_M$ from moiré disorder [39]. Disorder is likely also responsible for the requirement of a small but finite magnetic field to induce the topological phase transition near $E_c$.

Compared to the original Haldane model, which concerns a spinless (or, equivalently, fully spin-polarized) system [1], the Haldane model realized here is a fully valley-polarized system. The Zeeman field lifts the spin-valley degeneracy. Band inversion and topological bands are present in only one of the valleys. The realization of the Haldane model is made possible by the electrically tunable band inversion and charge gap at $\nu = 2$ in AB-stacked MoTe$_2$/WSe$_2$ moiré bilayers.

**Methods**

**Device fabrication**

We fabricated the dual-gated $MoTe_2/WSe_2$ devices using the layer-by-layer dry transfer technique [7,29,40]. The device constituents, including graphite, hBN, $MoTe_2$ and $WSe_2$ crystals, were exfoliated onto Si substrates with a thermally grown $SiO_2$ layer. $MoTe_2$ crystals were handled in a glovebox with $O_2$ and $H_2O$ concentrations below 1 part per million (ppm) to prevent sample degradation. We first fabricated the bottom gate by picking up a hBN flake (thickness ~ 10 nm) and a few-layer graphite flake and depositing them on a Si substrate with pre-patterned Ti/Au electrodes. After dissolving the polymer residual, we deposited 5-nm Pt contacts on hBN by standard electron-beam lithography and evaporation. This is followed by another step of electron beam lithography and metallization to form 5-nm Ti/40-nm Au to connect the thin Pt contacts on hBN to pre-patterned electrodes. We cleaned the Pt contacts and the hBN surface after lift-off using the atomic force microscope (AFM) contact mode. Future transfers were performed in the glovebox. In the devices, the $MoTe_2$ and $WSe_2$ monolayers are aligned at 60°. This was achieved by determining the crystal orientations of the monolayers using the optical second-harmonic generation technique before stacking [41,42]. The angle alignment was verified using the same technique. We chose a relatively thin hBN layer (~ 4 nm) for the top gate to achieve large breakdown electric fields (~ 1 V/nm). We also chose a narrower top graphite gate electrode than the bottom gate, which defines the region of interest.

**Electrical measurements**

Electrical measurements were carried out in a closed-cycle $^4$He cryostat (Oxford TeslatronPT) with magnetic fields up to 14 T and temperatures down to 300 mK (using a $^3$He insert) and in a Bluefors LD250 dilution refrigerator with magnetic fields up to 12 T and lattice temperatures down to 10 mK. The standard low-frequency (10 - 20 Hz) lock-in technique was used to measure the sample resistance under low bias (0.2 - 1 mV) to avoid sample heating. A voltage pre-amplifier with 100-M$\Omega$ impedance was used to measure the sample resistance up to 10 M$\Omega$. Longitudinal and transverse voltage drops and source-drain current were recorded. Finite longitudinal-transverse coupling occurs in our devices. We used the standard procedure to obtain the longitudinal and Hall resistance by symmetrizing $\frac{R_{xx}(B) + R_{xx}(-B)}{2}$ and anti-symmetrizing $\frac{R_{xy}(B) - R_{xy}(-B)}{2}$ the measured $R_{xx}$ and $R_{xy}$ values under positive and negative magnetic fields, respectively.

**Estimate of the Stark shift**

The shift of the band offset between the $MoTe_2$ and $WSe_2$ monolayers in the TMD heterobilayer under a perpendicular electric field $E$ is evaluated as $DE$, where $D$ is the electric dipole moment of the TMD heterobilayer, and $E = (\frac{V_{bg}}{d_{bg}} - \frac{V_{tg}}{d_{tg}})/2$ is defined as the average field right above and below the TMD heterobilayer. Here $V_{bg}$ and $V_{tg}$ are the bottom and top gate voltages, and $d_{bg}$ and $d_{tg}$ are the thickness of the bottom and top hBN gate dielectrics, respectively. The field $E$ is related to the field in the TMD heterobilayer, $E_{TMD}$, by continuity of the displacement field in the device, $\varepsilon_{hBN}E = \varepsilon_{TMD}E_{TMD}$, where $\varepsilon_{hBN} \sim 3$ and $\varepsilon_{TMD} \sim 7$ are the out-of-plane dielectric constant of hBN and the TMD heterobilayer, respectively [43,44]. The Stark shift can also be expressed as $E_{TMD} \, et$, where $t \sim 0.7 \, nm$ is the interlayer separation between the two TMD layers. We thus derive the electric dipole moment $D = \frac{\varepsilon_{hBN}}{\varepsilon_{TMD}} et \sim 0.26 \, e \times nm$, which is in good agreement with the value from optical measurements [29].

**Tight-binding model**

The moiré potential minima lie at the MM stacking region in the MoTe$_2$ layer and the XX region in the WSe$_2$ layer in the 60°-aligned bilayers (M = Mo, W; X = Te, Se) [11]. These potential minima form a staggered honeycomb lattice with the sublattice potential tunable by the out-of-plane electric field $E$. Because of the 60°-alignment, the lowest-energy valence bands in two layers have opposite spins in the same valley. The spin-conserving nearest-neighbor hopping is strongly suppressed due to the large momentum mismatch between the band edges of the two valance bands. The nearest-neighbor hopping is a spin-flip process; it is allowed by the broken mirror symmetry in the z direction. We adopt the tight-binding Hamiltonian of Ref. [14], $H = H_A + H_B + H_{AB}$, to model the lowest-energy valence moiré bands under an external out-of-plane magnetic field. Here $H_A$ and $H_B$ are the tight-binding Hamiltonian of each sublattice (layer), and $H_{AB}$ is the hybridization or inter-layer hopping term:

$$H_\alpha = -\sum_{<i,j>_\alpha,\sigma} \left( t_\alpha \, e^{is_\sigma v_{i,j}\phi_\alpha} a_{i\sigma}^\dagger a_{j\sigma} + h.c. \right) - \sum_{i\in\alpha} \frac{\tau_\alpha \Delta}{2} n_i + \sum_{i\in\alpha,\sigma} V_z s_\sigma n_{i\sigma},$$

$$H_{AB} = -t_{AB} \sum_{<i,j>,\sigma} a_{i\sigma}^\dagger a_{j\bar\sigma}.$$

Here $a^\dagger$, $a$ and $n$ are the electron creation, annihilation, and number operators, respectively; $\alpha = A, B$ denotes the two sublattices (layers) with $\tau_A = -\tau_B = 1$; $\sigma = \uparrow, \downarrow$ denotes the z-component electron spin with $s_\uparrow = -s_\downarrow = 1$; the next-nearest-neighbor (or intralayer layer) hopping is assumed to have equal amplitude $t_A = t_B = t_2$ and same phases for simplicity; $v_{i,j} = \pm 1$ denotes the two opposite hopping directions; $\Delta$ is the layer potential difference; and $V_z$ is the Zeeman energy.

We note that in the special case of $\phi_A = \phi_B = \pi/2$ the above Hamiltonian reduces to the original Kane-Mele model [16], which is practically two time-reversal copies of the Haldane model. In the original Haldane model (for spinful particles), the band inversion occurs between two orbitals of the same spin indices, $|A,\uparrow\rangle$ and $|B,\uparrow\rangle$. Here the Bloch Hamiltonian remains the same as the phase of the next-nearest-neighbor hopping changing sign twice upon a 60° lattice rotation and a spin flip; the band inversion occurs between two orbitals with different spin indices [14,45], $|A,\uparrow\rangle$ and $|B,\downarrow\rangle$.

Extended Data Fig. 6 shows the first two moiré valence bands under different electric and magnetic fields. Hopping parameters, $t_{AB} = 1$ meV, $t_2 = 2$ meV, $\phi_A = \phi_B = 2\pi/3$ have been used. The direct gap size (between the two moiré bands at the K and K' point of the moiré mini Brillouin zone) is shown as a function of the interlayer potential difference $\Delta$ and Zeeman energy $V_z$. No global band gap is observed. The tight binding approximation is known to underestimate the gap size. The Hartree-Fock approximations, which incorporate the electron-electron interactions, have been shown to yield a global band gap [12]. Our result shows that in the absence of a magnetic field, the out-of-plane electric field induces a phase transition from a QSH insulator (after band inversion) to a normal band insulator (before band inversion) at the critical point $\Delta_c \sim 9$ meV. When the interlayer potential is fixed slightly above $\Delta_c$ (before band inversion), a magnetic field induces a transition from a normal band insulator to a Haldane Chern insulator, and the critical field increases linearly with $(\Delta - \Delta_c)$. Similarly, a magnetic-field-induced transition is observed from a QSH insulator to a Haldane Chern insulator after band inversion.

**Landau fans and band inversion**

We study the magneto-transport characteristics under high magnetic fields to better understand the effect of the Landau levels. Extended Data Figure 7 shows the result from device 2 at 10 mK (lattice temperature). Panel a is the gate voltage dependences of $R_{xy}$ (top) and $R_{xx}$ (bottom) under a fixed magnetic field of 11.8 T. Panel b and c are the magnetic-field and filling dependences of $R_{xy}$ and $R_{xx}$ under two out-of-plane electric fields, corresponding to before (b) and near band inversion (c), respectively. Before band inversion, a set of Landau fans originated from $\nu = 2$ starts to emerge under about 3 T. The onset field provides an estimate of the disorder-induced energy level broadening ($\sim 0.5$ meV). Clear Landau fans can be observed only for $\nu > 2$. This is presumably associated with doping into the WSe$_2$ layer, which has a substantially higher carrier mobility than in the MoTe$_2$ layer. The Landau level degeneracy is determined to be 1. This is also consistent with the known valley- and spin-polarized Landau levels in hole-doped WSe$_2$ (Ref. [46,47]).

Near band inversion, a state with Chern number $c = 1$ emerges under low magnetic fields (below 3 T). This is the magnetic-field-induced Haldane insulator state discussed in the main text. In contrast, here we observe Landau levels with both positive and negative slopes; there are also multiple level crossings. The presence of Landau levels with opposite dispersion and yet the same sign for $R_{xy}$ (all corresponding to hole transport) supports the inverted band structure. Similar effect has been observed in HgTe [23] and InAs/GaSb quantum wells [48]. The energy of the Landau levels near the inverted band edge and away from it disperses with magnetic field in opposite directions [49].

**Acknowledgement**
The authors thank Liang Fu, Yang Zhang and Allan MacDonald for fruitful discussions.

**Author contributions**
W.Z. and L.L fabricated the devices. W.Z., K.K. L.L, C.T. and E.R. performed the measurements and analyzed the data. K.K. performed the tight binding model calculations. K.W. and T.T. grew the bulk hBN crystals. W.Z., J.S. and K.F.M. designed the scientific objectives and oversaw the project. All authors discussed the results and commented on the manuscript.

**Figures**

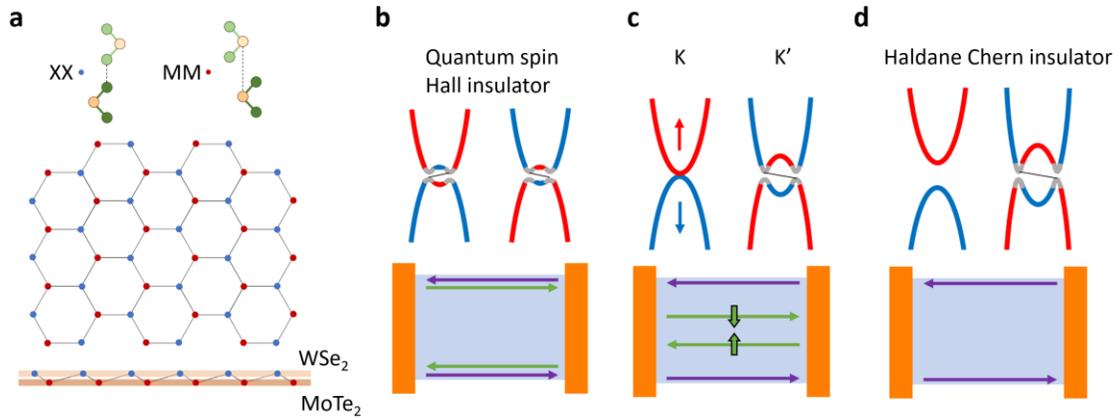

**Figure 1 | AB-stacked MoTe$_2$/WSe$_2$ moiré bilayer. a**, Top: atomic stacking structure at the high-symmetry MM and XX sites (orange balls for M = Mo or W; green balls for X = Te or Se). Middle: a staggered honeycomb moiré lattice formed by the MM (red) and XX (blue) sites. Bottom: side view. The electron Wannier wavefunction at the MM (XX) site is centered in the MoTe$_2$ (WSe$_2$) layer. **b-d**, Top: schematics of the electronic band structure. Red and blue lines denote spin-up and spin-down bands. The black lines represent the edge states. Bottom: schematics of the edge states in a stripe of samples (light blue). Contacts are shown in orange. The bands are inverted at both the K and K′ valleys with helical edge states bound to the sample edges (**b**). This corresponds to a quantum spin Hall insulator. Upon the application of a critical magnetic field, the bands in the K valley cross at a single point; the corresponding edge state is pushed deep into the bulk (**c**). With further increase of the magnetic field, the gap in the K valley is positive; a Chern insulator with a chiral edge state is formed (**d**).

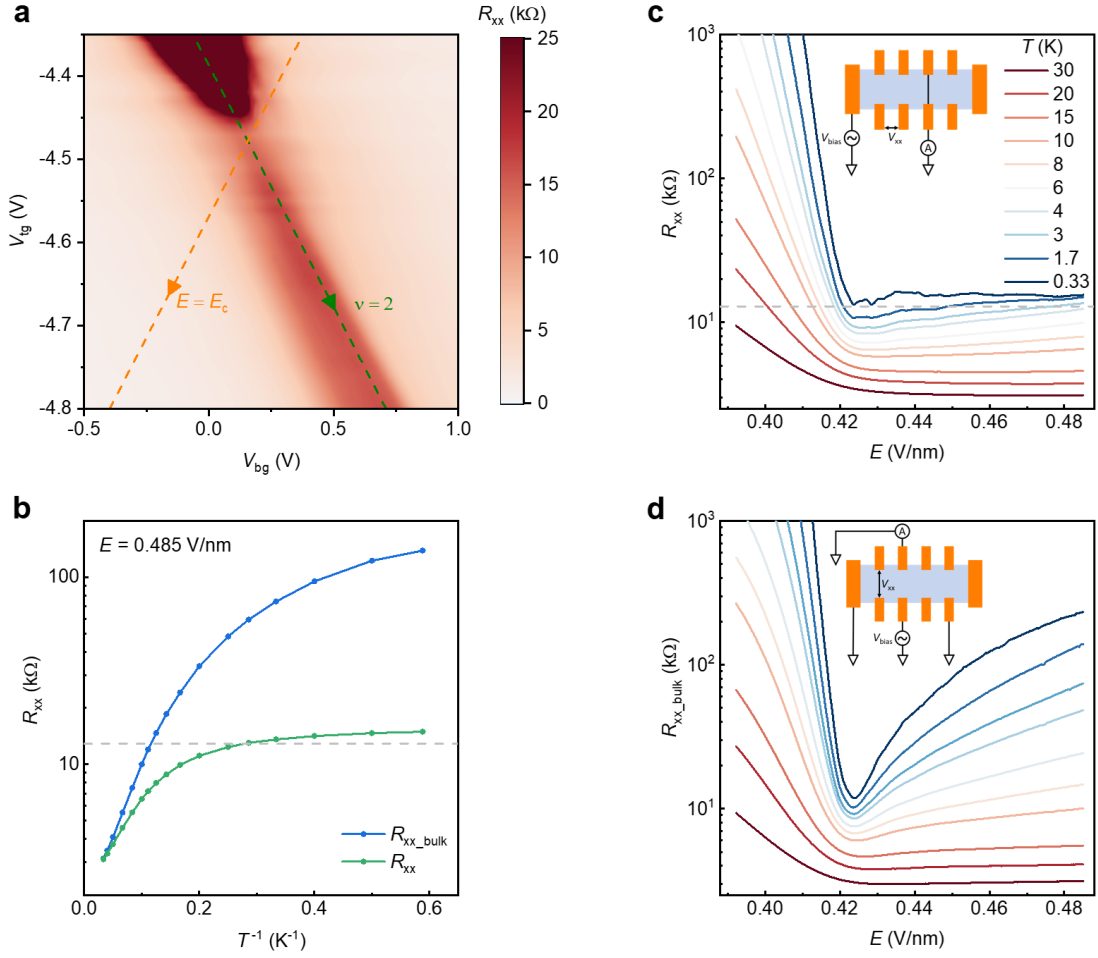

**Figure 2 | Quantum spin Hall insulator at $\nu = 2$. a,** Dependence of $R_{xx}$ on the top and bottom gate voltages. The green and orange dashed lines denote, respectively, the electric field direction at fixed $\nu = 2$ and the filling factor direction at fixed $E = E_c$. **b,** Arrhenius plot for the edge and bulk resistance, $R_{xx}$ and $R_{xx\_bulk}$, at $E = 0.485\ V/\text{nm}$ (above $E_c$). The symbols are the measurement result and the lines are guides to the eye. To account for the difference in the measurement geometry, $R_{xx\_bulk}$ is multiplied by a factor of 1.3 to match $R_{xx}$ in the high-temperature limit. At low temperatures, $R_{xx}$ is nearly quantized at $\left(\frac{1}{2}\right)\frac{h}{e^2}$ (dashed line). **c, d,** Electric-field dependence of $R_{xx}$ (**c**) and $R_{xx\_bulk}$ (**d**) at varying temperatures. The insets show the measurement configurations. To measure $R_{xx\_bulk}$, two contacts are grounded to avoid the edge current.

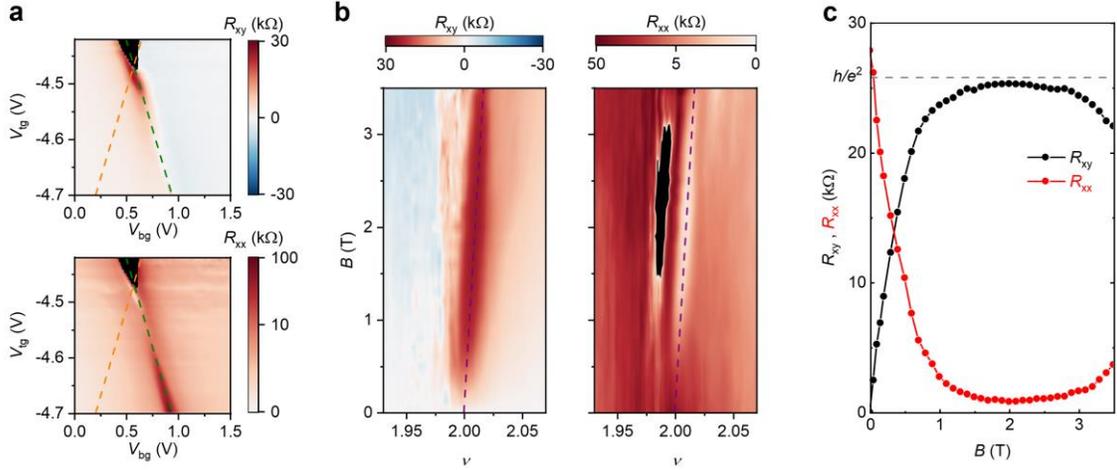

**Figure 3 | Magnetic-field-induced Haldane Chern insulator. a,** Hall resistance (top) and longitudinal resistance (bottom) at 10 mK (lattice temperature) and $B = 2$ T as a function of the top and bottom gate voltages. **b,** Magnetic field and filling factor dependences of $R_{xy}$ (left) and $R_{xx}$ (right) at $E = E_c$ (along the orange dashed line in **a**). The filling factor for the $R_{xy}$ maximum and the $R_{xx}$ minimum disperses linearly with magnetic field (dashed lines). The best-fit slope is $n_M \frac{d\nu}{dB} = (1.00 \pm 0.05) \frac{e}{h}$. Transport characteristics cannot be reliably probed in the black regions due to poor electrical contacts. **c,** Magnetic-field dependence of $R_{xx}$ and $R_{xy}$ along the dashed lines in **b**. The lines are guides to the eye. The Hall resistance is nearly quantized at $\frac{h}{e^2}$ (dashed line) between 1 T and 3 T.

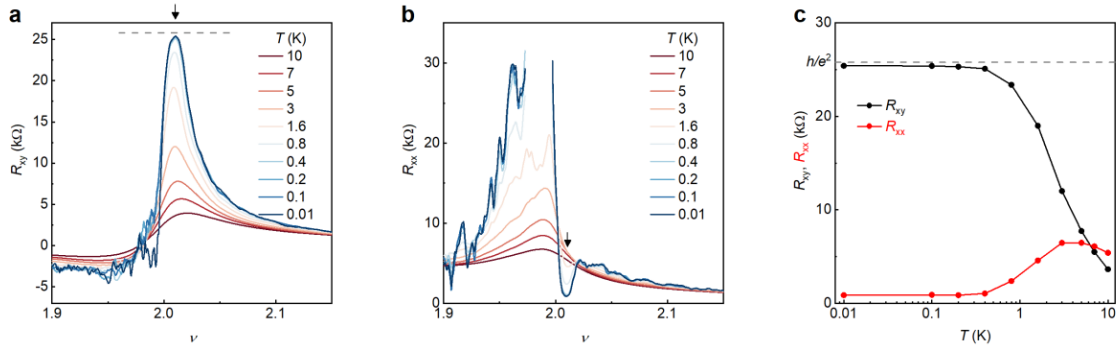

**Figure 4 | Temperature dependence of the Chern insulator. a, b,** Filling factor dependence of $R_{xy}$ (**a**) and $R_{xx}$ (**b**) at $B = 2$ T and $E = E_c$ for different temperatures. The arrows mark the emergence of the $R_{xy}$ peak, which is approximately quantized at $\frac{h}{e^2}$ (dashed line) and the $R_{xx}$ dip, which is below 1 kΩ at low temperatures. **c,** Temperature dependence of $R_{xx}$ and $R_{xy}$ of the Chern insulating state. The dashed line marks $\frac{h}{e^2}$. The solid lines are a guide to the eye.

**Extended data figures**

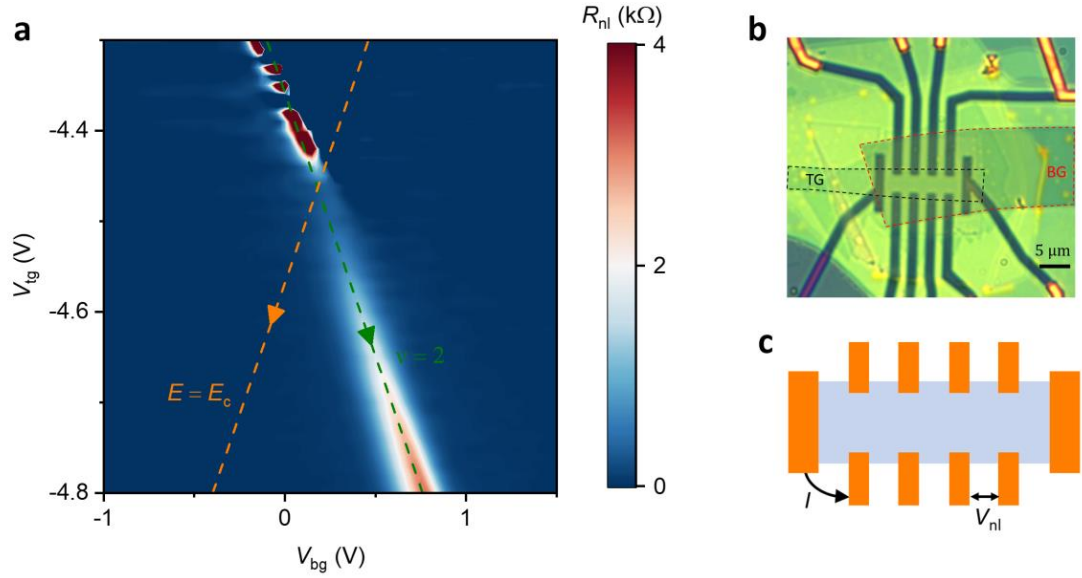

**Extended Data Figure 1 | Nonlocal transport supporting the quantum spin Hall insulator. a,** Nonlocal resistance $R_{nl}$ as a function of top and bottom gate voltages at $B = 0$ and $T = 330$ mK. The green and orange dashed lines denote, respectively, the electric field direction at fixed $\nu = 2$ and the filling factor direction at fixed $E = E_c$. **b,** Optical micrograph of device 1. The top gate (TG) and bottom gate (BG) are outlined by a black and red dashed line, respectively. The scale bar is 5 µm. **c,** Schematics of the measurement geometry. The device is current ($I$) biased along the arrow direction. The voltage drop $V_{nl}$ is measured at the other end of the device. $R_{nl}$ is negligible except at $\nu = 2$. Before band inversion, $R_{nl}$ cannot be probed appropriately because the current is practically zero; after band inversion, it grows with increasing electric field. This is consistent with the helical edge transport in a quantum spin Hall insulator.

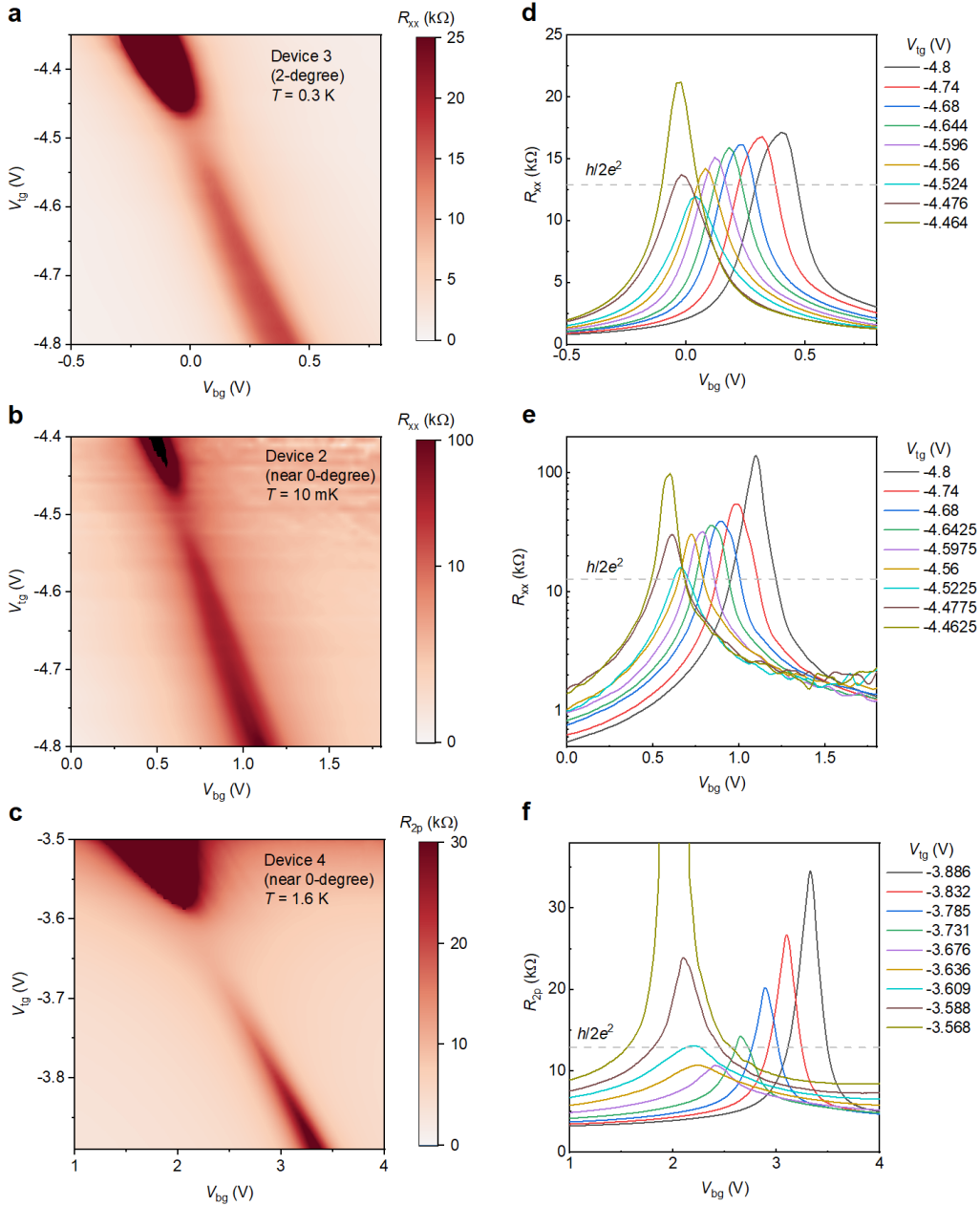

**Extended Data Figure 2 | Quantum spin Hall effect in devices of different twist angles.**
**a, b,** Longitudinal resistance as a function of top and bottom gate voltages of device 3 (2-degree twisted) at $T = 300$ mK (**a**) and device 2 (near 0-degree twisted) at $T = 10$ mK (**b**).
**c,** Two-terminal resistance measured by adjacent contacts as a function of top and bottom gate voltages of a bilayer $MoTe_2$/monolayer $WSe_2$ device 4 (near 0-degree twisted) at $T = 1.6$ K. **d-f,** Line-cuts of **a-c** at varying top gate voltages. The grey dashed lines mark the quantized resistance $\frac{h}{2e^2}$ ($\approx 12.9$ k$\Omega$). For device 3, $R_{xx}$ at $\nu = 2$ is the smallest right after band inversion and plateaus with further increase of the electric field. $R_{xx}$ and $R_{2p}$ increase continuously with electric field after band inversion for device 2 and 4, respectively.

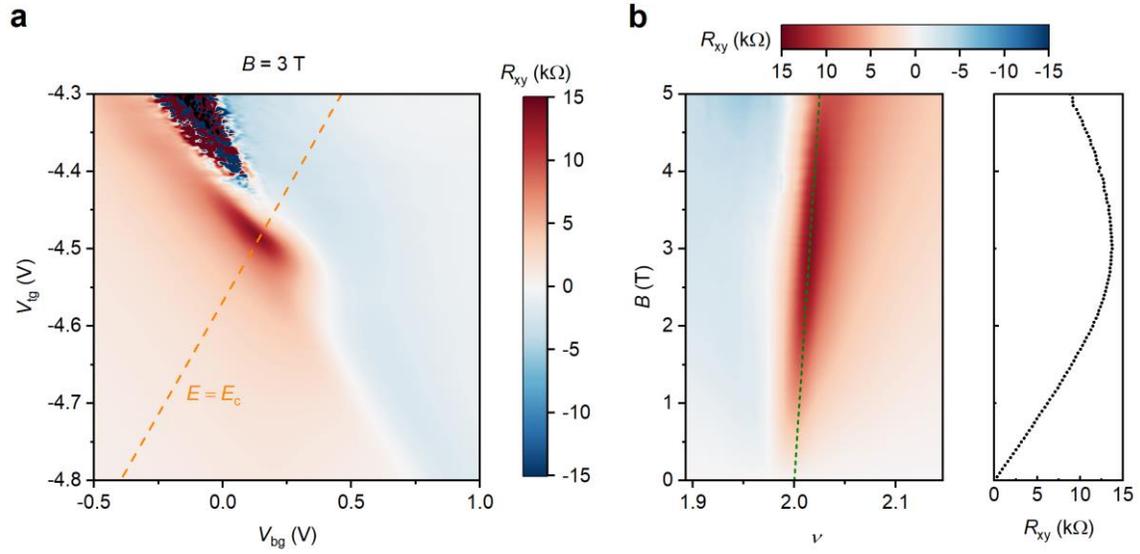

**Extended Data Figure 3 | Magneto-transport in device 1 at 1.6 K. a, b,** Hall resistance as a function of the top and bottom gate voltages at $B = 3$ T (**a**) and of the magnetic field and filling factor at $E = E_c$ (**b**). The orange dashed line in **a** corresponds to $E = E_c$. The green dashed line in **b** marks the $R_{xy}$ maximum. We determine the slope of the dashed lines to be $n_M \frac{d\nu}{dB} = c\frac{e}{h}$ with $c = 1.1 \pm 0.1$. Right inset shows the magnetic-field dependence of $R_{xy}$ along the green dashed line.

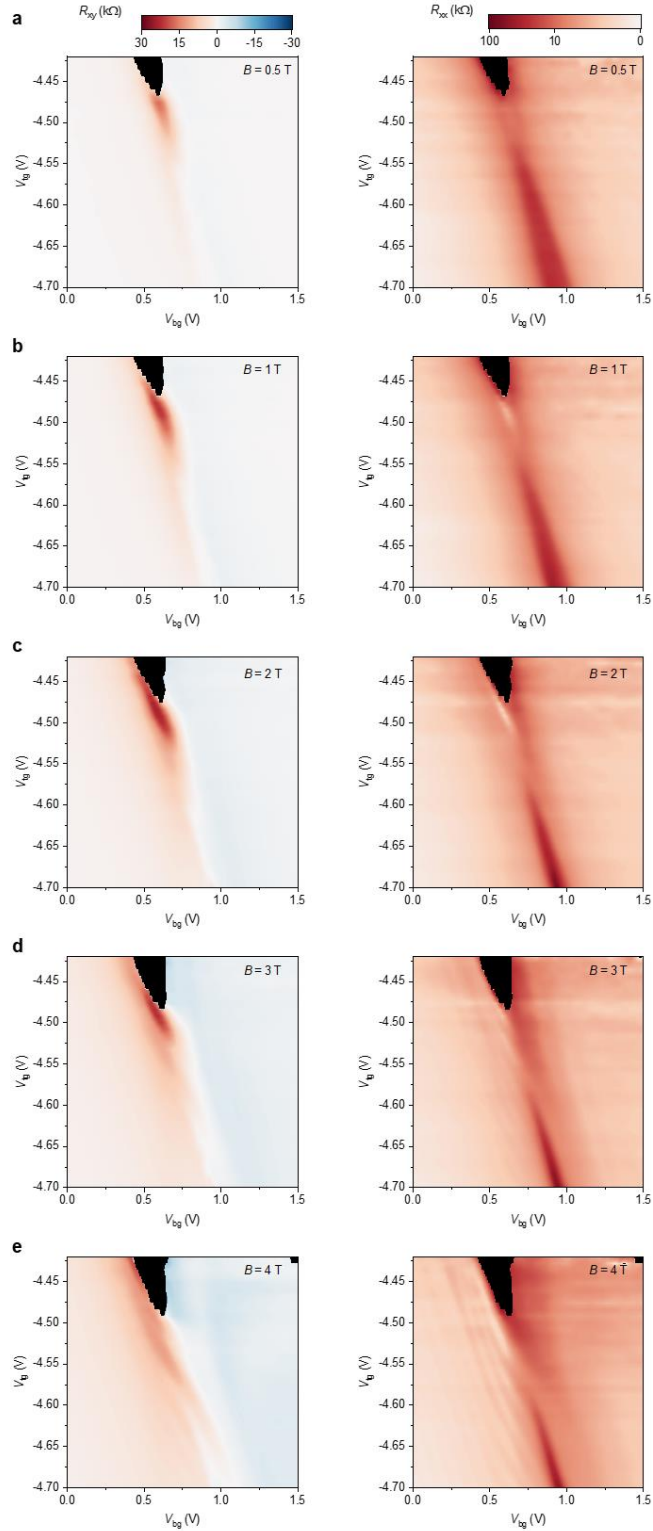

**Extended Data Figure 4 | Transport under different magnetic fields for device 2. a-e**, Hall resistance (left) and longitudinal resistance (right) at 10 mK (lattice temperature) as a function of the top and bottom gate voltages. The magnetic field is 0.5 T (**a**), 1 T (**b**), 2 T (**c**, same as Fig. 3a of the main text), 3 T (**d**), 4 T (**e**).

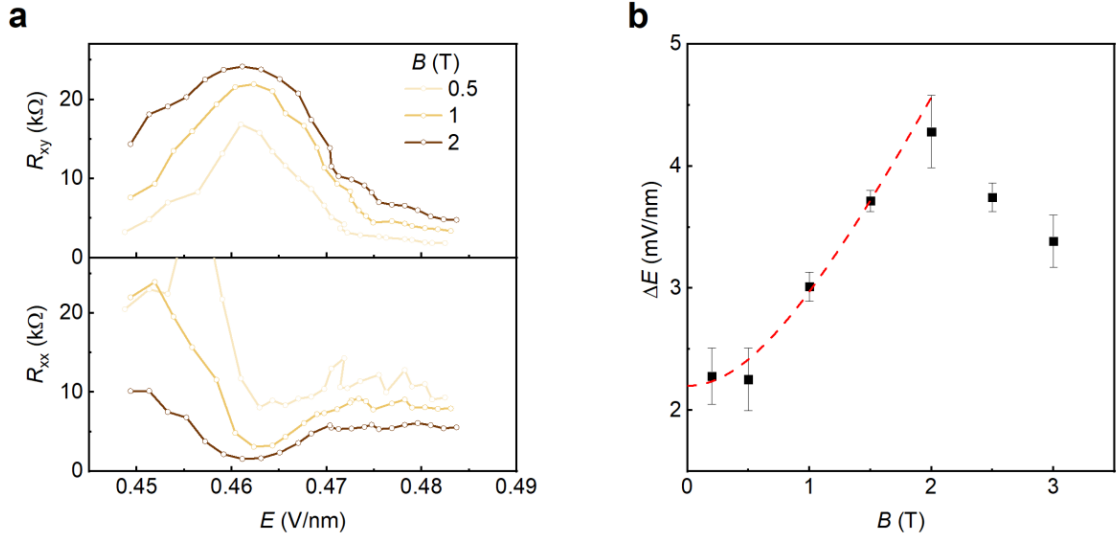

**Extended Data Figure 5 | The electric-field span of the Chern state as a function of magnetic field. a**, $R_{xy}$ and $R_{xx}$ of device 2 at $\nu = 2$ and $T = 10$ mK (lattice temperature) as a function of electric field. The data is extracted from Extended Data Fig. 4. **b**, The electric-field span, $\Delta E$, of the Chern state as a function of magnetic field. For each magnetic field, the electric-field dependence of $R_{xy}$ is fit with a Gaussian function and $\Delta E$ is determined as half of the variance. The red dashed line is the best fit to $\Delta E = \sqrt{a^2 + (bB)^2}$ with fitting parameters $a = 2.2 \pm 0.1$ mV/nm and $b = 2 \pm 0.1$ mV/nmT.

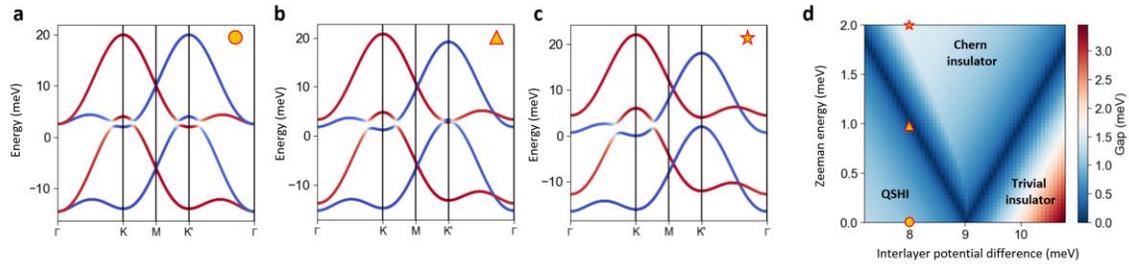

**Extended Data Figure 6 | Tight-binding model calculations. a-c**, Band structure simulated using the tight-binding Hamiltonian described in Methods. Red and blue curves denote the spin-up and spin-down bands, respectively. Under zero magnetic field (**a**), the bands are inverted at both the K and K′ valleys. This is a quantum spin Hall insulator (QSHI). Under a small magnetic field (**b**), the bands in the K′ valley cross at one momentum. Under a sufficiently high magnetic field (**c**), the gap changes sign for the K′ valley; this is a Chern insulator. **d**, The direct band gap near the K/K′ valleys as a function of the Zeeman energy and the sublattice/interlayer potential difference. The three symbols mark the phase space for which the electronic band structure is represented in **a-c**, respectively.

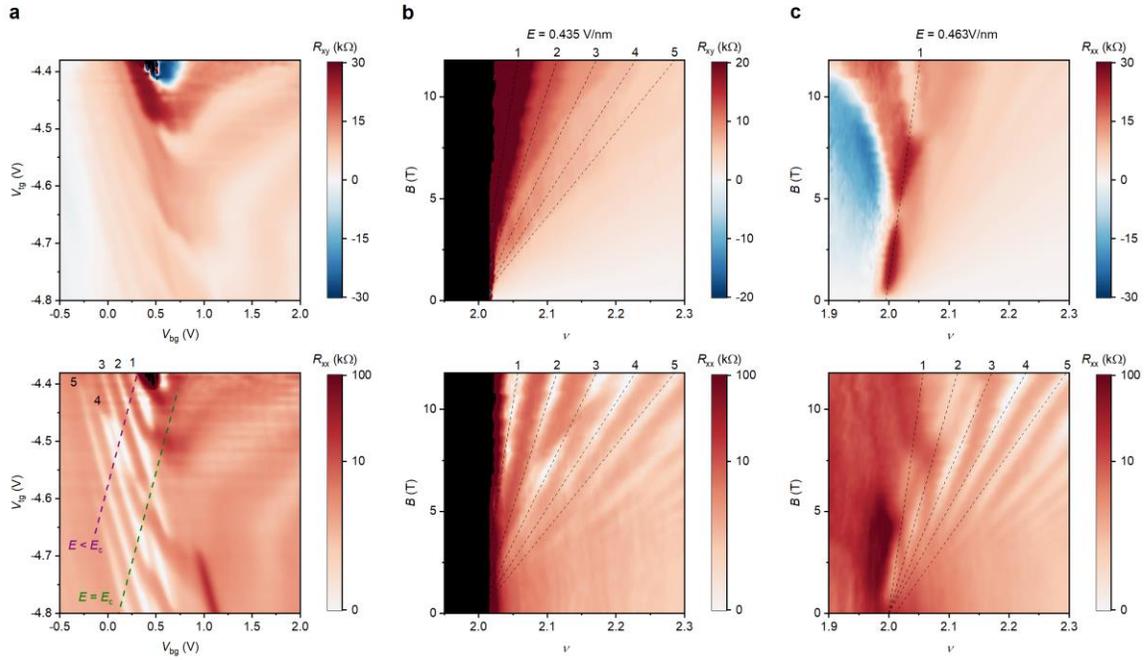

**Extended Data Figure 7 | High-magnetic-field transport. a,** $R_{xy}$ (top) and $R_{xx}$ (bottom) of device 2 as a function of top and bottom gate voltages at $B = 11.8$ T and $T = 10$ mK (lattice temperature). The Landau levels (identified by the resistance minimum) are indexed. **b, c,** $R_{xy}$ (top) and $R_{xx}$ (bottom) as a function of filling factor and magnetic field at $E = 0.435$ V/nm (before band inversion, **b**) and at $E = 0.463$ V/nm (after band inversion, **c**). The dashed lines show the Landau fan.